\documentclass[twocolumn,trackchanges]{aastex631}
 \usepackage{natbib}
\bibpunct[; ]{(}{)}{;}{a}{}{,}
\def\daox{$\Delta\alpha_{\rm ox}$}
\def\simlt{\lower 2pt \hbox{$\, \buildrel {\scriptstyle <}\over {\scriptstyle \sim}\,$}}

\newcommand{\simgt}{\lower 2pt \hbox{$\, \buildrel {\scriptstyle >}\over {\scriptstyle \sim}\,$}}
\newcommand{\xray}{\hbox{\it \rm \hbox{X-ray}\/}}
\newcommand{\aox}{$\alpha_{\rm OX}$}
\let\oldAA\AA
\renewcommand{\AA}{\text{\normalfont\oldAA}}

\usepackage{subfigure} 
\begin{document}
\title{The Remarkable X-ray Spectra and Variability of the Ultraluminous Weak-Line Quasar SDSS J1521+5202}
\author[0009-0003-1260-4143]{Shouyi Wang}
\affiliation{School of Astronomy and Space Science, Nanjing University, Nanjing, Jiangsu 210093, People’s Republic of China}
\affiliation{Key Laboratory of Modern Astronomy and Astrophysics (Nanjing University), Ministry of Education, Nanjing 210093, People’s Republic of China}
\affiliation{Department of Astronomy and Astrophysics, 525 Davey Lab, The Pennsylvania State University, University Park, PA 16802, USA}

\author[0000-0002-0167-2453]{W. N. Brandt}
\affiliation{Department of Astronomy and Astrophysics, 525 Davey Lab, The Pennsylvania State University, University Park, PA 16802, USA}
\affiliation{Institute for Gravitation and the Cosmos, The Pennsylvania State University, University Park, PA 16802, USA}
\affiliation{Department of Physics, 104 Davey Lab, The Pennsylvania State University, University Park, PA 16802, USA}

\author[0000-0002-9036-0063]{Bin Luo}
\affiliation{School of Astronomy and Space Science, Nanjing University, Nanjing, Jiangsu 210093, People’s Republic of China}
\affiliation{Key Laboratory of Modern Astronomy and Astrophysics (Nanjing University), Ministry of Education, Nanjing 210093, People’s Republic of China}

\author[0000-0002-6990-9058]{Zhibo Yu}
\affiliation{Department of Astronomy and Astrophysics, 525 Davey Lab, The Pennsylvania State University, University Park, PA 16802, USA}
\affiliation{Institute for Gravitation and the Cosmos, The Pennsylvania State University, University Park, PA 16802, USA}

\author[0000-0002-4436-6923]{Fan Zou}
\affiliation{Department of Astronomy and Astrophysics, 525 Davey Lab, The Pennsylvania State University, University Park, PA 16802, USA}
\affiliation{Institute for Gravitation and the Cosmos, The Pennsylvania State University, University Park, PA 16802, USA}

\author[0000-0002-9335-9455]{Jian Huang}
\affiliation{School of Astronomy and Space Science, Nanjing University, Nanjing, Jiangsu 210093, People’s Republic of China}
\affiliation{Key Laboratory of Modern Astronomy and Astrophysics (Nanjing University), Ministry of Education, Nanjing 210093, People’s Republic of China}

\author[0000-0002-8577-2717]{Qingling Ni}
\affiliation{Max-Planck-Institut f\"{u}r extraterrestrische Physik (MPE), Gie{\ss}enbachstra{\ss}e 1, D-85748 Garching bei M\"unchen, Germany}

\author[0000-0003-0680-9305]{Fabio Vito}
\affiliation{INAF -- Osservatorio di Astrofisica e Scienza dello Spazio di Bologna, Via Gobetti 93/3, I-40129 Bologna, Italy}
\begin{abstract}
We present a focused \hbox{X-ray} and multiwavelength study of the
ultraluminous weak-line quasar (WLQ) SDSS J1521+5202, one of the few \hbox{X-ray} weak WLQs that is
amenable to basic \hbox{X-ray} spectral and variability investigations. J1521+5202 shows striking \hbox{X-ray}
variability during 2006--2023, by up to a factor of $\approx 32$ in 0.5--2~keV flux, and our new 2023 Chandra observation caught it in its brightest \hbox{X-ray} flux state to date. Concurrent infrared/optical observations show only mild variability. The 2023 Chandra spectrum can be acceptably described by a power law with intrinsic \hbox{X-ray} absorption, and it reveals a nominal intrinsic level of \hbox{X-ray} emission relative to its optical/ultraviolet emission. In contrast, an earlier Chandra spectrum from 2013 shows apparent spectral complexity that is not well fit by a variety of models, including ionized-absorption or standard Compton-reflection models. Overall, the observations are consistent with the thick-disk plus outflow model previously advanced for WLQs, where a nominal level of underlying \hbox{X-ray} emission plus variable absorption lead to the
remarkable observed \hbox{X-ray} variability. In the case of J1521+5202 it appears likely that the outflow, and not the thick disk itself, lies along our line-of-sight and causes the \hbox{X-ray} absorption.
\end{abstract}
\section{Introduction}

\subsection{Observations and Modeling of Weak-Line Quasars} \label{sec:intro_1}

Weak-line quasars are a sub-class of type~1 quasars that continue to provide
insights into quasar accretion physics and structure. They are blue, luminous quasars with remarkably weak and often highly blueshifted high-ionization emission lines. For example, their broad C~IV rest-frame equivalent widths (REWs) are \hbox{$\simlt 10$--15~\AA}, and their C~IV blueshifts can
reach \hbox{5000--10000~km~s$^{-1}$} \citep[e.g.,][]{Fan+1999,Diamond+2009,Wu+2012}. Most WLQs are radio quiet. 

X-ray observations of WLQs have played an essential role in revealing
their nature \citep[e.g.,][]{Wu+2011,Luo+2015,marlar+2018,Ni+2018,Ni+2022}. WLQs show a number of remarkable \xray\ properties, as recently detailed in Section~1 of \citet{Ni+2022}. Briefly, WLQs show an unusually broad range of \xray\ luminosities compared to those expected from their optical/ultraviolet (UV) luminosities
or spectral energy distributions, with about half of WLQs being notably \xray\ weak. The \xray\ luminosity relative to the optical/UV luminosity is generally assessed using the \aox\ and \daox\ parameters. \aox\ is the slope of a nominal power law 
connecting the rest-frame 2500~\AA\ and 2~keV monochromatic luminosities, i.e., 
$\alpha_{\rm ox}=0.3838 \log(L_{\rm 2~keV}/L_{2500~\mathring{\rm{A}}})$, and this quantity is significantly correlated with $L_{2500~\mathring{\rm{A}}}$ \citep[e.g.,][]{steffen+2006,Just+2007}. We also define 
$\Delta\alpha_{\rm ox}=\alpha_{\rm ox}({\rm Observed})-\alpha_{\rm ox}(L_{2500~\mathring{\rm{A}}})$, which quantifies the deviation of the observed \xray\ 
luminosity relative to that expected from the \aox--$L_{2500~\mathring{\rm{A}}}$ correlation. \daox\ can be used to derive factors of \xray\ weakness following 
$f_{\rm weak}=403^{-\Delta\alpha_{\rm ox}}$.
\xray\ spectral analyses of the half of WLQs with nominal-strength
(relative to their optical/UV emission given by the \aox--$L_{2500~\mathring{\rm{A}}}$ relation) \xray\ emission show steep power-law continua (with photon indices of
\hbox{$\Gamma=2.0$--2.4}), suggesting high Eddington ratios ($L$/$L_{\rm Edd}$). 
In contrast, \xray\ spectral analyses of the half of WLQs that are
\xray\ weak show hard \xray\ spectra on average (effective \hbox{$\langle \Gamma\rangle\approx 1.2$--1.4}), suggesting high levels of intrinsic \xray\ absorption
(at least $N_{\rm H}\approx 10^{23}$~cm$^{-2}$) and probably also Compton reflection. Such heavy \xray\ absorption is surprising
given these quasars’ type 1 nature and blue optical/UV continua without
Broad Absorption Lines (BALs) or other strong UV absorption features.

Based upon the available \xray\ and multiwavelength results, a basic working model
has been advanced for WLQs that has the potential to explain, in a simple
and unified manner, their weak UV lines, their \xray\ properties, and their other
multiwavelength properties \citep[e.g.,][]{Wu+2011,Luo+2015,Ni+2018,Ni+2022}; Figure~1 of \citet{Ni+2018} shows a relevant schematic of the model. 
To explain the weak UV lines, this model relies upon
small-scale ``shielding'' of ionizing EUV/\xray\ photons that prevents
them from reaching the broad emission-line region (BELR).
The shielding material is likely the geometrically and optically thick inner
accretion disk, and its associated outflow, expected for a quasar
accreting with high $L$/$L_{\rm Edd}$---the thick disk and its outflow
will be abbreviated as ``TDO''. The shielding is also
responsible for the \xray\ weakness and apparent absorption seen in about
half of WLQs. When our line of sight intercepts the shield, we see an \xray\
weak WLQ; when it misses the shield, we observe an \xray\ normal WLQ.
In both cases, ionizing EUV/\xray\ photons are prevented from reaching 
the (largely equatorial and unobscured) high-ionization BELR, and 
the optical/UV continuum remains unobscured. 
In this shielding model, strong \xray\ flux and spectral variability could
arise if the line-of-sight absorption column density and/or covering factor of the shield varies, e.g., due to motions of the TDO. Indeed, a few examples of
remarkably strong \xray\ variability discovered among WLQs have been explained
with TDO variations \citep[e.g.,][]{Minuutti+2012,2020Ni,Liu+2022}. In extreme cases, the observed \xray\ flux could vary between \xray\ weak (highly absorbed) and \xray\ normal (completely
unabsorbed) states. 

\subsection{The Ultraluminous Weak-Line Quasar J1521+5202}

X-ray spectral and variability analyses of the critical \xray\ weak WLQs are challenging,
owing to the inevitably limited numbers of detected counts (generally 10 or
fewer) in the available observations \citep[e.g.,][]{Ni+2022}.
Most of the \xray\ spectral results for these objects have been derived
only in an average sense by stacking sets of WLQs together,
which can confuse interpretation if spectral diversity is present among
the objects stacked.

Moderate-quality \hbox{X-ray} spectral analyses have only been possible for the singular
object J1521+5202, an \xray\ weak WLQ at $z=2.24$ that is one of the few most-luminous quasars in
the Universe in the optical/UV with $M_i=-30.2$ \citep{2005+Schneider}.  
The exceptional luminosity of \hbox{J1521+5202} makes it
possible to obtain \xray\ spectra despite its clear \xray\ weakness
by a factor of $f_{\rm weak}\approx 35$
\citep[$\alpha_{\rm ox}=-2.42$, $\Delta\alpha_{\rm ox}=-0.59$, and 
observed \hbox{$L_{\rm 2-10~keV}=3\times 10^{44}$~erg~s$^{-1}$};][]{Luo+2015}. The rest-frame optical/UV spectroscopic properties of \hbox{J1521+5202} are representative of those of WLQs. For example, its C~IV
line is strikingly weak ($\mathrm{REW}=9.1\pm 0.6$~\AA) but is clearly broad (FWHM $=11700\pm 800$~km~s$^{-1}$) and blueshifted (by $9300\pm 610$~km~s$^{-1}$).
Based on (admittedly uncertain) virial estimates using its H$\beta$ line, its
black-hole mass is $\approx 6\times 10^9$~M$_\odot$ with 
\hbox{$L/L_{\rm Edd}\approx 1$--2} \citep{Wu+2011,Luo+2015}. 
Similar to the other \xray\ weak WLQs, \hbox{J1521+5202} does not show any BALs
in its UV spectrum. It is a radio-quiet quasar with $R<0.2$, where 
$R=f_{\rm 5\;GHz}/f_{4400\mbox{\rm~\scriptsize\AA}}$. 

A Chandra Advanced CCD Imaging Spectrometer (ACIS) spectrum of \hbox{J1521+5202}
(37.2~ks exposure; 92 counts), obtained in October 2013, revealed
extraordinary properties for a highly luminous, type~1, and non-BAL quasar \citep{Luo+2015}.
Fitting a power-law model with Galactic absorption (using the Cash statistic) returned a photon index of $\Gamma=0.6\pm 0.2$ for the rest-frame \hbox{3--20~keV} band. This very hard \xray\ spectrum suggests the \xray\ weakness of \hbox{J1521+5202} is likely due to strong intrinsic \xray\ absorption. 

\subsection{Paper Overview and Definitions}

In this paper, we present the most complete \xray\ spectral and variability study
of \hbox{J1521+5202} to date, aiming to understand better the nature of this remarkable
WLQ and WLQs generally. We have obtained a new Chandra ACIS observation
(29.7~ks exposure), and we also collect all available archival \xray\
results for this quasar over the past 18~yr to examine its long-term \xray\ variability. We furthermore compare with its long-term optical and infrared (IR) variability. In Section~2 we present the \xray\ observations utilized and their reduction, and in Section~3 we present the derived \xray\ and multiwavelength properties of \hbox{J1521+5202}. Section~4 presents a summary of the results with relevant discussion. 

Throughout this paper, we use J2000 coordinates and a cosmology
with $H_0=70$~km~s$^{-1}$~Mpc$^{-1}$, $\Omega_\Lambda=0.7$, and 
$\Omega_M=0.3$. 
\section{X-ray Observations and Data Reduction}\label{sec:X_obs}
\subsection{Chandra Observations}\label{sec:Chan_obs}
J1521+5202 was observed by Chandra \hbox{ACIS-S} three times in June 2006, October 2013, and February 2023. Table~\ref{tab:obslog} provides the basic information for these Chandra observations. The results from the first and second observations were originally presented in \citet{Just+2007} and \citet{Luo+2015}, respectively, and we present results from the third observation here for the first time. 
\par
Using the Chandra Interactive Analysis of Observations (CIAO; v4.15) tools, we analyze the Chandra data. We first run the {\sc chandra\_repro} script for each observation to generate a new level 2 event file. Then, we run the {\sc deflare} script and use an iterative $3\sigma$ clipping algorithm to filter potential background flares. The cleaned exposure times are 4.1~ks, 37.2~ks, and 29.7~ks in 2006, 2013, and 2023, respectively. We use the {\sc specextract} tool to extract source spectra using a circular source region with a radius of 4\arcsec\ centered on the quasar \hbox{X-ray} position. The background spectra are extracted from an annular region centered on the quasar position with a 6\arcsec\ inner radius and a 15\arcsec\ outer radius. We group the spectra with at least one count per bin for spectral fitting. The spectra have $\approx 3$, 92, and 114 total counts in the full band (\hbox{0.5--8~keV}) in 2006, 2013, and 2023, respectively. We then assess the source-detection significance via computing the binomial no-source probability \citep[e.g.,][]{Luo+2015,Liu+2022}, $P_\mathrm{B}$, which is defined as
 \begin{equation} \label{equ1}
     P_{\mathrm{B}}=\sum_{X=S}^{N} \frac{N!}{X!(N-X)!} p^{X}(1-p)^{N-X}.
 \end{equation}
 In Equation (\ref{equ1}), $S$ is the total number of counts in the source region in the full band, $B$ is the total number of counts in the background region, $N = S + B$, and $p$ = 1 /(1 + BACKSCAL), with BACKSCAL being the ratio of the background and source region areas. The computed $P_\mathrm{B}$ values in these three Chandra observations are all smaller than 0.01, corresponding to $\textgreater 2.6~\sigma$ detection significance levels. Therefore, J1521+5202 is considered to be detected in all Chandra observations. Note that productive spectral analyses are possible with the second and third observations despite the limited numbers of counts, since the signal-to-noise ratio is high with only \hbox{3--4} background counts expected in the source cell (see Table~\ref{tab:obslog}).
\subsection{XMM-Newton Observations}\label{sec:XMM_obs}
Two XMM-Newton observations of J1521+5202 taken in July 2019 and December 2019 are listed in Table~\ref{tab:obslog}. The total observation times are 81~ks in July and 80~ks in December. The \hbox{X-ray} data were processed using the \hbox{XMM-Newton}  Science Analysis System \citep[SAS;][]{SAS2004}. We followed the standard procedure in the SAS Data Analysis Threads.\footnote{\url{https://www.cosmos.esa.int/web/xmm-newton/sas-threads}.} Background flares were filtered to generate cleaned event files. For each detector, source and background spectra were extracted with a \hbox{10\arcsec-radius} circular source region and a \hbox{50\arcsec-radius} circular \hbox{source-free} background region on the same CCD chip. Spectral response files were generated using the tasks {\sc rmfgen} and {\sc arfgen}. For the July observation, the effective exposures in each detector are 42.1~ks in pn, 45.0~ks in MOS1, and 65.1~ks in MOS2. In (pn, MOS1, MOS2), we find (57, 31, 36) total counts (source+background) and (22.8, 10.4, 18.3) background-subtracted counts in the \hbox{0.5--10~keV} energy range, and the total number of source counts in the three detectors is 51.5. For the December observation, the effective exposures in each detector are 46.6~ks in PN, 57.8~ks in MOS1, and 72.0~ks in MOS2. In (pn, MOS1, MOS2), we find (82, 38, 24) total counts (source+background) and (22.6, 24.0, 12.4) background-subtracted counts in the \hbox{0.5--10~keV} energy range, and the total number of source counts in the three detectors is 59.0.
\par
 J1521+5202 was in a low-flux state during both XMM-Newton observations, and the resulting data have limited signal-to-noise ratios. Therefore, we will not attempt detailed spectral analyses of the XMM-Newton data beyond simple power-law spectral fitting.

\begin{deluxetable*}{lccccc}
\tabcolsep=0.28cm
\tablecaption{\hbox{X-ray} observation log}
\tablehead{
\colhead{Observatory}  &
\colhead{Observation}  &
\colhead{Observation} &
\colhead{Exposure Time}&
\colhead{Net counts}&
\colhead{Background counts}
\\
\colhead{ }   &
\colhead{ID}   &
\colhead{Start Date}   &
\colhead{(ks)}&
\colhead{}&
\colhead{}
}
\decimalcolnumbers
\startdata
Chandra&6808&2006-07-16&4.1&2.8&0.3\\
  &15334&2013-10-22&37.2&88.4&3.6\\
  &27364&2023-02-02 &29.7&111.0&3.1\\
  XMM-Newton&0840440101&2019-07-26&42.1 (pn)&22.8 (pn)& 34.2(pn)\\
  &0840440201&2019-12-16&46.6 (pn)&22.6 (pn)& 59.4 (pn)
\enddata
\tablecomments{Column (1): Name of the \xray\ observatory. Column (2): Observation ID. Column (3): Observation start date. Column (4): Cleaned exposure time. Column (5): Background-subtracted counts. Column (6): Background counts in the source cell.}
\label{tab:obslog}
\end{deluxetable*}
\section{X-ray and Multiwavelength Properties}\label{sec:propt}
\subsection{Basic X-ray Spectral Analyses}\label{sec:X_analy}
\begin{figure*}[htbp]
  \centering
   \subfigure {\includegraphics[width=0.48\textwidth,height=2.7in]{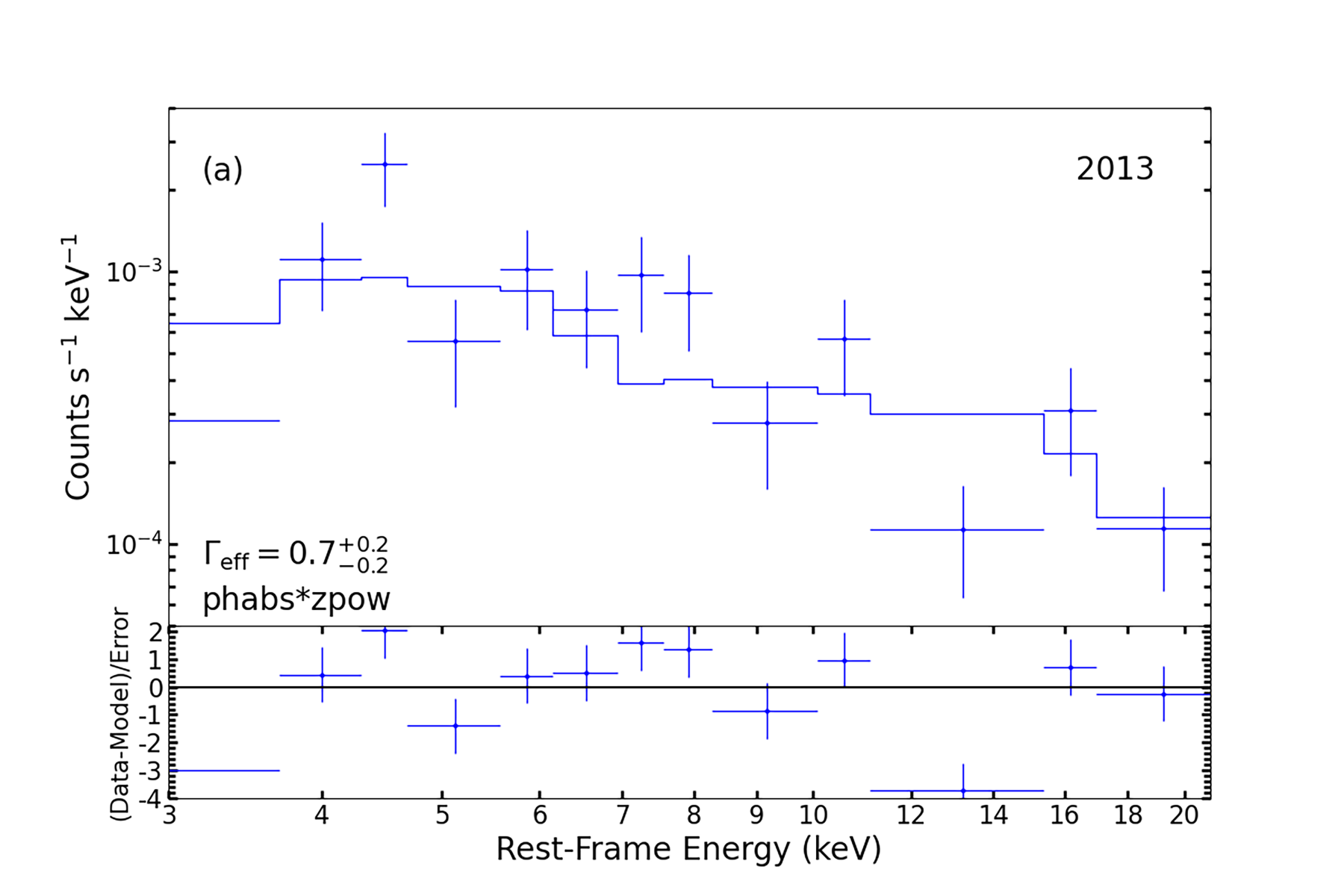}\label{fig:figure1a}}
   \quad
  \subfigure {\includegraphics[width=0.48\textwidth,height=2.7in]{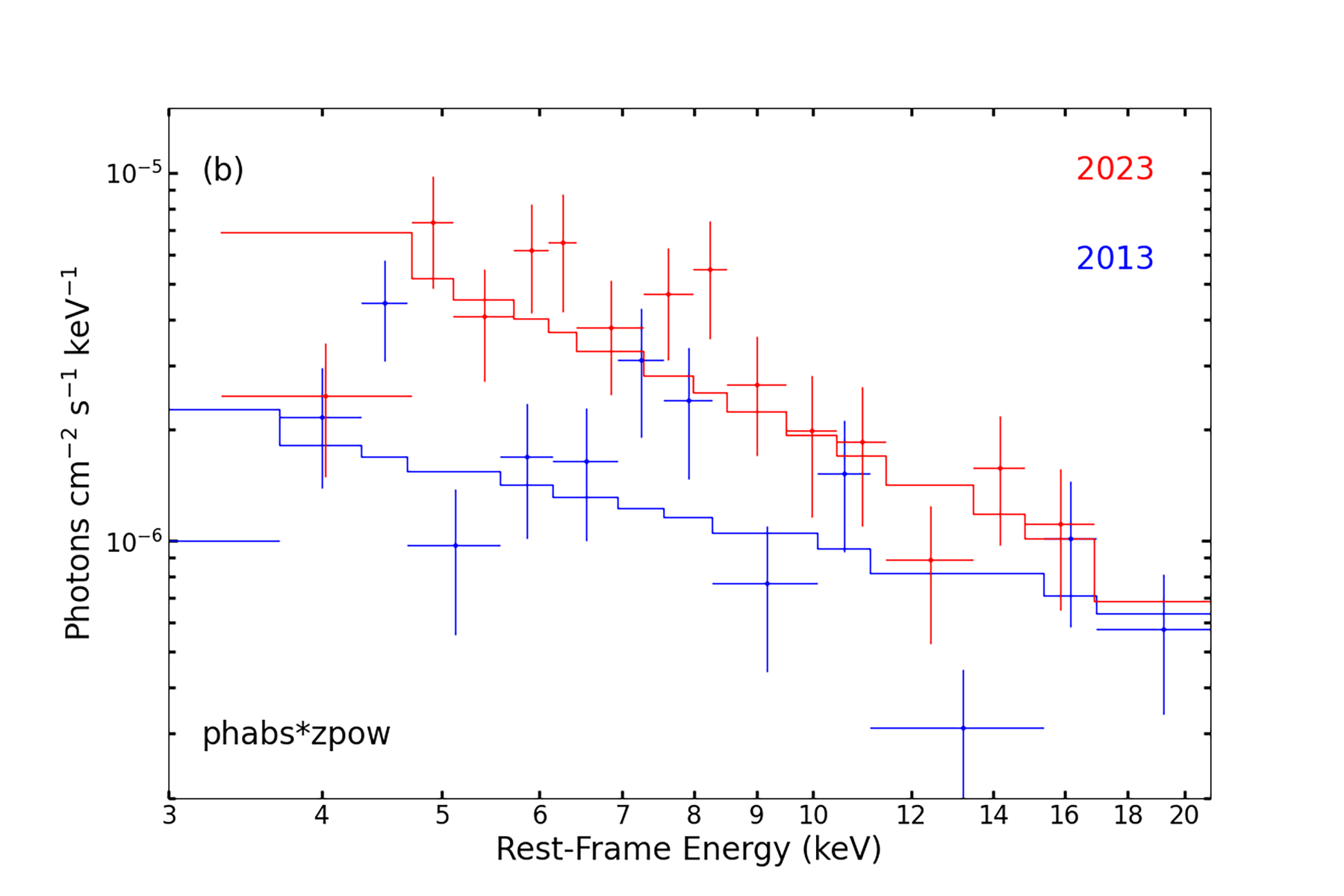}\label{fig:figure1b}}
  \quad
  \subfigure {\includegraphics[width=0.48\textwidth,height=2.7in]{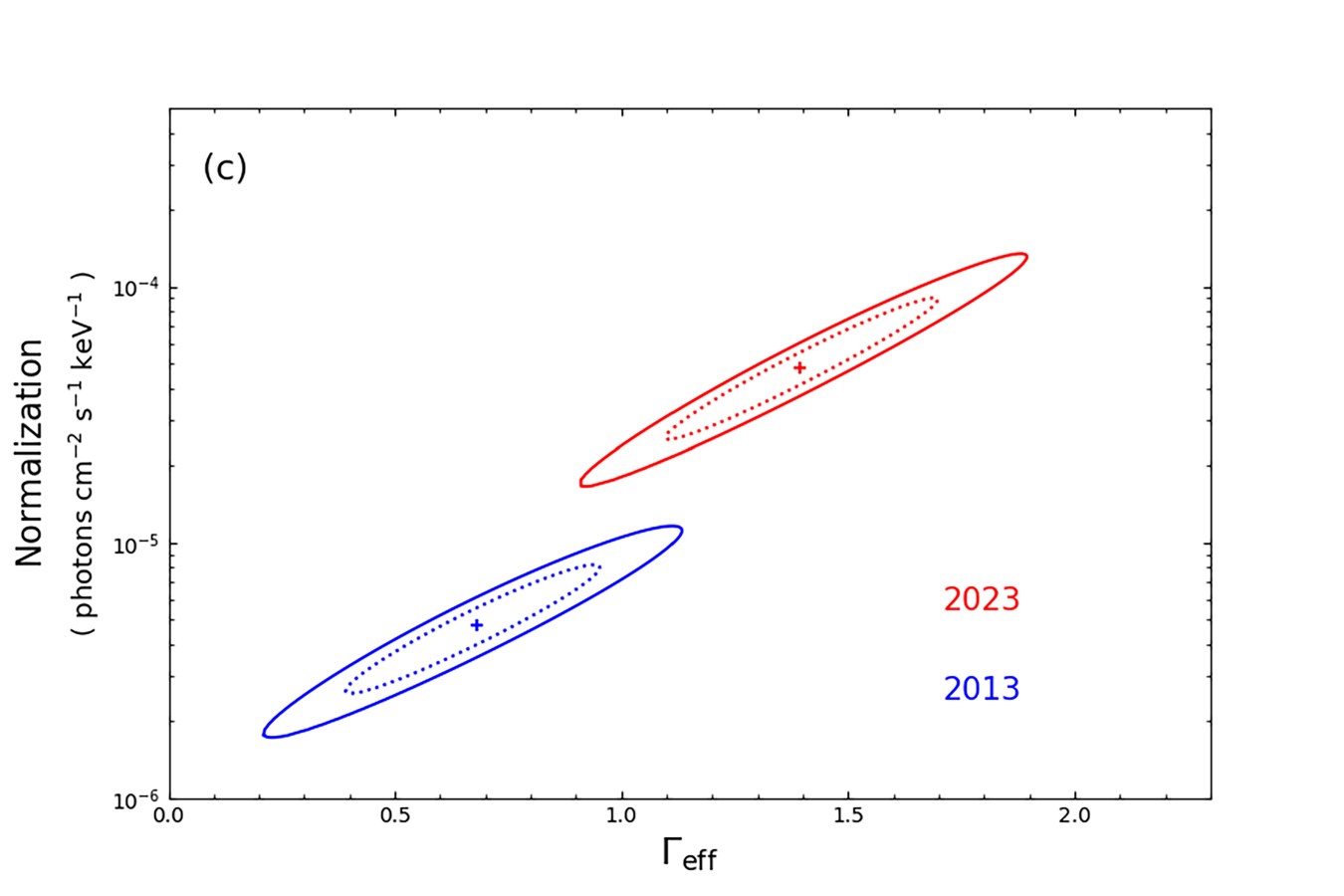}\label{fig:figure1c}}
  \quad  
   \subfigure {\includegraphics[width=0.48\textwidth,height=2.7in]{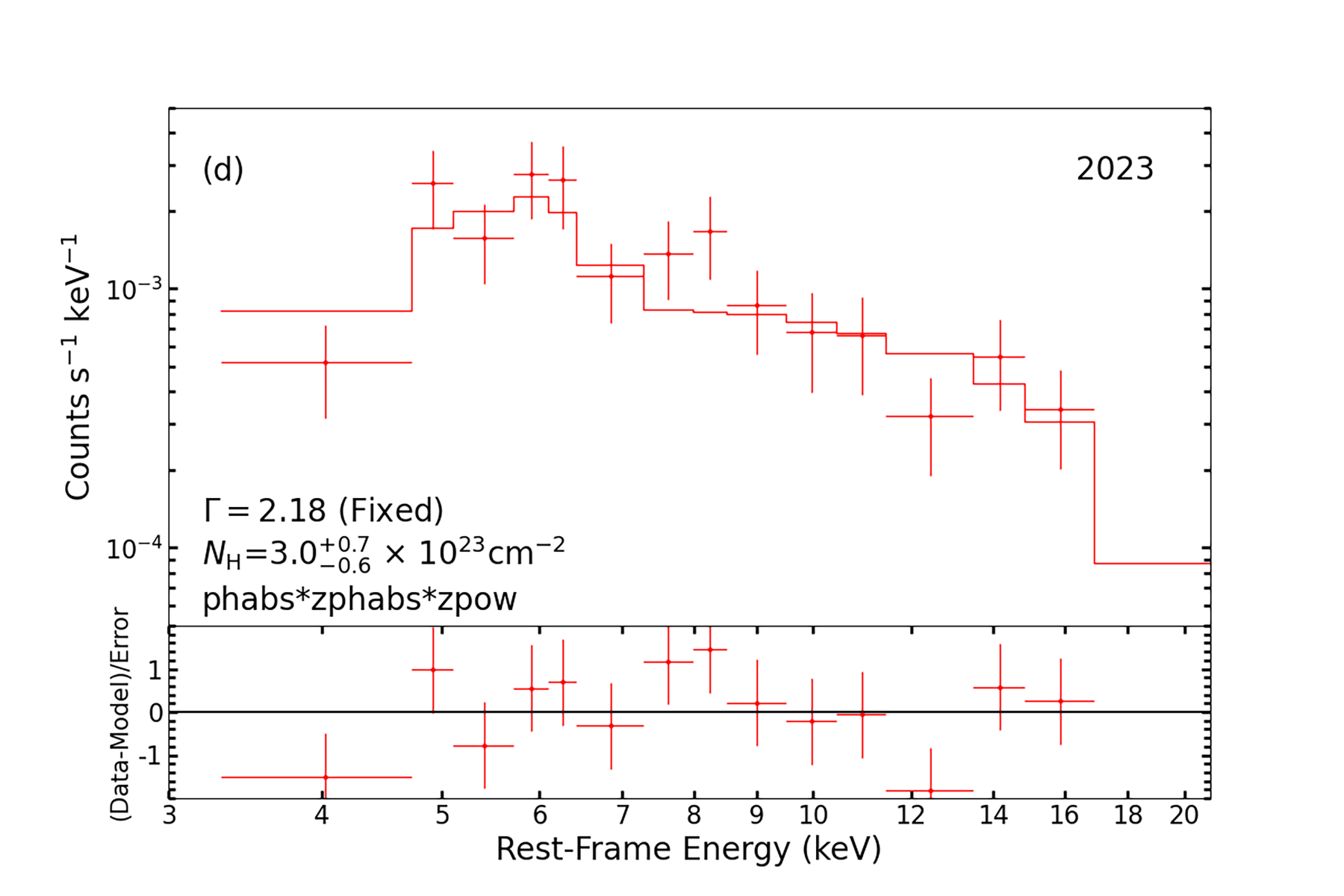}\label{fig:figure1d}}
   \quad
  \caption{(a): Chandra 2013 spectrum overlaid with the best-fit simple power-law model. The bottom panel displays the fitting residuals. The data are grouped for display purposes only. (b): Unfolded Chandra 2013 (blue) and 2023 (red) spectra overlaid with the best-fit simple power-law models, showing the spectral shape and flux variations between the two spectra. (c): The $\Gamma_\mathrm{eff}$ and normalization values (blue and red plus symbols) and their 1$\sigma$ (dotted) and 2$\sigma$ (solid) contours for the best-fit simple power-law models.  (d): Chandra 2023 spectrum overlaid with the best-fit absorbed power-law model. The bottom panel displays the fitting residuals.} \label{fig:nomal}
\end{figure*}

The rest-frame $\approx 3-20$ keV Chandra and XMM-Newton spectra were fitted using \texttt{sherpa}  \citep{2001sherpa,2007sherpa}. The \textit{W}-statistic was used due to the limited photon counts. The Galactic neutral hydrogen column density was fixed at $1.58 \times 10^{20}~\mathrm{cm}^{-2} $ \citep{HI4PICollaboration}. To start, we adopted a simple power-law model modified by Galactic absorption to describe the 0.5--8 keV (for Chandra) or 0.5--10 keV (for XMM-Newton) spectra. The model was \texttt{phabs*zpowerlw}, where \texttt{zpowerlw} modeled the \xray~continuum, and \texttt{phabs} described the Galactic absorption. The effective power-law photon index derived from this modeling, $\Gamma_\mathrm{eff}$, serves to characterize the basic spectral shape. We adopted a Monte Carlo approach to assess the goodness-of-fit \citep[e.g.,][]{2017+kaastra}. The distribution of \hbox{fit-statistic} values is obtained by running simulations of the spectrum 10000 times using the best-fit model, instrument response, and exposure time. We also include parameter scatter in the simulations, where each input parameter set is one sampled realization of the fitting of the actual data. The fraction of simulated fit-statistic values smaller than our best-fit statistic value ($P_\mathrm{rej}$) represents the confidence level that the model can be rejected. The best-fit results are summarized in Table~\ref{tab_obs}, and the best-fit model for the 2013 Chandra spectrum is shown in Figure~\ref{fig:figure1a}. We also calculated $\alpha_\mathrm{OX}$ using the rest-frame 2 keV flux density ($f_\mathrm{2\, \rm keV}$) and rest-frame \hbox{2500~\AA}~flux density ($f_{2500 \, \textup{\AA}}$). The latter was converted from the $r$-band light curve (see Section ~\ref{sec:SEDlc} below). The $\Delta \alpha_\mathrm{OX}$ and $f_\mathrm{weak}$ values are also listed in \hbox{Table~\ref{tab_obs}}, and J1521+5202 showed remarkable \hbox{X-ray} weakness ($f_\mathrm{weak}$ = 4.5--298). 

There appears to be substantial spectral variability of J1521+5202. The 2006 Chandra and 2019 XMM-Newton spectra have limited signal-to-noise ratios, and thus we focus the comparison on the 2013 and 2023 Chandra spectra. The large $P_\mathrm{rej}$ value for the 2013 spectrum (Table~\ref{tab_obs}) suggests that the simple power-law model is inadequate to describe the data. However, given the difficulties in finding a better spectral model (see Section~\ref{sec:testmodels} below), we still consider that the power-law model provides a basic overall description of the spectral shape (via $\Gamma_\mathrm{eff}$) and flux level (via the normalization parameter) of the spectrum. In Figures~\ref{fig:figure1b} and~\ref{fig:figure1c}, we show the unfolded best-fit power-law spectra and the 1--2$\sigma$ contours of the $\Gamma_\mathrm{eff}$ and power-law normalization parameters, respectively, demonstrating spectral variability between these two observations.

The small $\Gamma_\mathrm{eff}$ values (Table~\ref{tab_obs}), compared to a typical value of $\Gamma$ = $2.18\pm0.09$ for \hbox{X-ray} normal WLQs \citep[][]{Luo+2015}, indicate that the significant \hbox{X-ray} weakness of this source is likely due to strong absorption with Compton reflection also possibly present. Therefore, we further added an intrinsic-absorption component as a next step to fit the 2013 and 2023 Chandra spectra. The full model in \texttt{sherpa} was \texttt{phabs*zphabs*zpowerlw}, where \texttt{zphabs} accounted for the intrinsic absorption. For the 2013 observation, the best-fit $\Gamma = 1.5_{-0.3}^{+0.4}$, and $N_\mathrm{H} = (1.3_{-0.5}^{+0.6})\times10^{23}$~cm$^{-2}$. For the 2023 observation, the best-fit $\Gamma = 3.1_{-0.5}^{+0.5}$, and $N_\mathrm{H} = (5.3_{-1.7}^{+1.4})~\times~10^{23}$~cm$^{-2}$ (Figure~\ref{fig:nomal}). The $P_\mathrm{rej}$ are 99.6\% and 27.9\% for the 2013 and 2023 fits, respectively. After considering intrinsic absorption, the two observations' best-fit $\Gamma$ values were still inconsistent with each other and the typical $\Gamma$ values of \hbox{X-ray} normal WLQs ($2.18\pm0.09$). Therefore, we further fixed the $\Gamma$ at 2.18 to assess what this implied for the intrinsic $N_\mathrm{H}$ and normalization. The best-fit $N_\mathrm{H}$ values were $(2.2_{-0.4}^{+0.4})~\times~10^{23}$~cm$^{-2}$ in the 2013 observation and $(3.0_{-0.6}^{+0.7})~\times~ 10^{23}$ cm$^{-2}$ in the 2023 observation. Besides, the normalization of the \texttt{zpowerlw} model increased by a factor of 2.5 from 2013 to 2023 with $\Gamma$ fixed at 2.18. The best-fit results are summarized in Table~\ref{tab_sim}, and the best-fit model for the 2023 Chandra spectrum is shown in Figure~\ref{fig:figure1d}. The $P_\mathrm{rej}$ are 99.8\% and 20.7\% for the 2013 and 2023 fits, respectively, and the high $P_\mathrm{rej}$ for 2013 suggests additional spectral complexity may be present. For the 2023 observation, we calculate the absorption-corrected $\alpha_\mathrm{OX,corr}=-1.80^{+0.02}_{-0.02}$ and $\Delta \alpha_\mathrm{OX,corr}=0.03$ with $\Gamma$ fixed at 2.18, indicating that J1521+5202 emitted a nominal intrinsic level of \hbox{X-ray} emission relative to its optical/UV emission at this epoch. 

 \begin{deluxetable*}{cccccccccc}
\tablecolumns{10}
\tabletypesize{\scriptsize}
\tabcolsep=0.25cm
\renewcommand{\arraystretch}{1.5}
\tablecaption{Best-fit results with a power-law model \label{tab_obs}}
\tablehead{
\colhead{Observation}&\colhead{$\Gamma_\mathrm{eff}$}&\colhead{log$F_\mathrm{X}$}&\colhead{$f_\mathrm{2\, \rm keV}$}&
\colhead{$f_{2500 \, \textup{\AA}}$}&\colhead{$\alpha_\mathrm{OX}$}&\colhead{$\Delta \alpha_\mathrm{OX}$}&\colhead{$f_\mathrm{weak}$}&\colhead{\textit{W}/\textit{d.o.f.}}&\colhead{$P_\mathrm{rej}$}\\
\colhead{ Time\small{[year]}}&\colhead{}&\colhead{(0.5--2 keV)}&\colhead{}&\colhead{}&\colhead{}&\colhead{}&\colhead{}&\colhead{}&\colhead{}}
\decimalcolnumbers
\startdata
Chandra [2006] & $-0.1^{+1.3}_{-1.0}$ &$-14.8^{+0.3}_{-0.4}$&$0.2^{+0.7}_{-0.2}$&24.4&$-2.70^{+0.5}_{-0.2}$&$-0.86$&$180^{+404}_{-165}$&0.04/1&12.9\%\\
Chandra [2013]&$0.7_{-0.2}^{+0.2}$&$-14.3_{-0.1}^{+0.1}$&1.2$_{-0.3}^{+0.4}$&22.3&$-2.40_{-0.04}^{+0.05}$&$-0.56$&$30^{+8}_{-7}$&104.6/76& 99.7\%\\
XMM-Newton [2019-07] & $0.4_{-0.4}^{+0.4}$ &$-15.2_{-0.2}^{+0.2}$ &0.11$_{-0.05}^{+0.08}$ &18.6&$-2.77_{-0.08}^{+0.12}$&$-0.95$&$298^{+184}_{-153}$&104.3/114&76.1\%\\
XMM-Newton [2019-12] & $0.9_{-0.4}^{+0.4}$ &$-15.1_{-0.2}^{+0.1}$ &0.19$_{-0.08}^{+0.12}$ &18.4&$-2.68_{-0.07}^{+0.11}$&$-0.86$&$195^{+107}_{-84}$&148.1/132&96.5\%\\
 Chandra [2023]& $1.4_{-0.2}^{+0.2}$& $-13.7_{-0.1}^{+0.1}$& 7.5$_{-2.6}^{+3.8}$& 19.4&$-2.08^{+0.08}_{-0.06}$&$-0.25$&$4.5^{+1.9}_{-1.7}$&80.7/85&66.2\%
\enddata
\tablecomments{Column (1): Observatory and observation year. Column (2): 0.5--8 keV effective power-law photon index. Column (3): Galactic absorption-corrected logarithm of observed-frame 0.5--2 keV flux. Column (4): Rest-frame \hbox{2 keV} flux density in units of $10^{-32}$~erg cm$^{-2} $s$^{-1}$Hz$^{-1}$. Column (5): Flux density at rest-frame 2500 \AA \; converted from the $r$-band light curve (Section~\ref{sec:SEDlc}) in units of $10^{-27}$~erg~cm$^{-2}$s$^{-1}$Hz$^{-1}$. Column (6): Measured $\alpha_\mathrm{OX}$ parameter.  Column (7): The difference between the measured $\alpha_\mathrm{OX}$ and the expected $\alpha_\mathrm{OX}$ from the $\alpha_{\rm OX}\textrm{--}L_{\rm 2500\AA}$ relation in \citet{Just+2007} . Column (8): Factor of \hbox{X-ray} weakness. Column (9): \textit{W}-statistic over degrees of freedom. Column (10): The fraction of simulated fit-statistic values smaller than our best-fit statistic value, representing the confidence level that the model can be rejected.}
\end{deluxetable*}

\begin{deluxetable*}{cccccc}
\tablecaption{ \hbox{Best-fit results with an absorbed power-law model}} \label{tab_sim}
\decimalcolnumbers
\renewcommand{\arraystretch}{1.5}
\tabcolsep=0.35cm
\tablehead{\colhead{Observation} & \colhead{$\Gamma$} & \colhead{$N_\mathrm{H}$ }&\colhead{Norm} &\colhead{\textit{W}/\textit{d.o.f.}}&\colhead{$P_\mathrm{rej}$}\\
 \colhead{year}&\colhead{}&\colhead{($10^{23}$ cm$^{-2}$)}&\colhead{}&\colhead{}&\colhead{}}
\startdata
Chandra [2013]&$1.5_{-0.3}^{+0.4}$& $ 1.3_{-0.5}^{+0.6} $& $0.4^{+0.5}_{-0.2}$& 95.5/75&99.6\%\\
Chandra [2023]& $3.1_{-0.5}^{+0.5}$ &$ 5.3_{-1.7}^{+1.4} $& $43^{+114}_{-30}$& 62.5/84&27.9\%\\
\hline
Chandra [2013]& 2.18 (fixed)&$2.2_{-0.4}^{+0.4}$&$1.8_{-0.3}^{+0.3}$ & 98.5/76&99.8\%\\
Chandra [2023]& 2.18 (fixed) &$3.0_{-0.6}^{+0.7}$&$4.4_{-0.6}^{+0.7}$&  66.2/85&20.7\%\\
\enddata
\tablecomments{ $\Gamma$ is a free parameter in the first two rows and is fixed at 2.18 in the last two rows. Column (1): Observatory and observation year. Column (2): Power-law photon index.  Column (3): Intrinsic absorption column density. Column (4): \hbox{Power-law} normalization in units of $10^{-4}$ photons $\rm cm^{-2} s^{-1} keV^{-1}$. Column (5): \textit{W}-statistic over degrees of freedom.  Column (6): The fraction of simulated fit-statistic values smaller than our best-fit statistic value, representing the confidence level that the model can be rejected.}
\end{deluxetable*}

\subsection{Testing of Additional X-ray Spectral Models} \label{sec:testmodels}
The relatively poor fit-quality results for the 2013 spectrum of J1521+5202
in Section \ref{sec:X_analy} suggest that additional spectral complexity was present
at this epoch. We have therefore attempted to fit the 2013 spectrum with
a variety of spectral models, including a partially covering absorption model, a \hbox{double-absorber} model, a broken power-law model with absorption, and Compton-thick absorption models.
Unfortunately, none of the models tested provides a statistically meaningful
improvement in fit quality as assessed using simulations;
e.g., the $P_\mathrm{rej}$ we obtain is 99.7\% for the \hbox{double-absorber} model. 
The lack of improvement in fit quality can be partly understood from the
lack of strong systematic residuals in Figure~\ref{fig:figure1a}, where the data points
appear generally scattered around the best fit but without a clear
systematic trend that can be straightforwardly modeled. 
There is likely additional spectral complexity present in 2013, but
with only 88 net counts, we are unable to constrain robustly the nature
of this complexity. Furthermore, the strong spectral variability between
2013 and 2023 makes it difficult to use the 2023 \hbox{X-ray} spectrum as a guide
for interpreting the 2013 spectrum better.

The strong \hbox{X-ray} weakness and hard \hbox{X-ray} spectral shape of J1521+5202 during
2013 suggest that heavy \hbox{X-ray} obscuration is present. It is therefore worth
testing if the 2013 spectrum is dominated by Compton reflection, which could
occur if the direct line-of-sight absorption column density toward the
nuclear \hbox{X-ray} source is Compton thick. If J1521+5202 were found to have
Compton-thick absorption, it would be among the most-luminous Compton-thick
quasars in the Universe with a bolometric luminosity comparable to the
powerful obscured AGNs in the most-luminous hot dust-obscured galaxies
\citep[hot DOGs; e.g., ][]{2018VITO}.

We have tested if the 2013 spectrum from rest-frame \hbox{3--20~keV} is
dominated by Compton reflection using the BORUS model for Compton reflection
from the obscuring tori of AGNs \citep{Borus+2018}.
Specifically, we tried the following model:
 \texttt{phabs*borus}, where \texttt{borus} is the reprocessed torus emission model of \citet{Borus+2018}. We fixed the $N_\mathrm{H}$ value of BORUS at $1~\times~10^{25}$~cm$^{-2}$, corresponding to highly Compton-thick material.
This model does not fit the data well; the $P_\mathrm{rej}$ is 99.99\%.
The primary reason this Compton-reflection dominated model fails to fit the data
is that the predicted \hbox{3--20~keV} continuum is even harder than what
is observed, owing to the rapid spectral rise toward the peak of the
Compton-reflection hump at 20--30 keV.

The BORUS model is derived for the case of Compton reflection from a
standard torus of AGN unified schemes, while the obscuring material for
WLQs is likely the TDO (see Section \ref{sec:intro_1}). The TDO is expected to be more
highly ionized than a standard torus, and it also will have rotational and/or outflow motions. Therefore, we have also considered the case
of Compton reflection from an ionized medium using the RELXILL model of \citet{2014D_rel} and \citet{2014Grel}, which incorporates XILLVER and RELCONV. The former calculates the reflection in the rest frame of an accretion disk, and the latter models the relativistic convolution. We tried the following model: \texttt{phabs*relxill}. We fixed the reflection fraction to $-1$ to model a reflection-dominated situation.\footnote{We assumed a power-law disk emissivity profile and fixed the dimensionless black-hole spin at 0.2 for simplicity. These parameters cannot be meaningfully constrained with our data.} The disk ionization parameter ($\xi$) is defined as $\xi=4\pi F_\mathrm{ion}/n$, in units of $\rm erg\,cm\,s^{-1}$, where $F_\mathrm{ion}$ is the ionizing irradiating flux, and $n$ is the number density of the disk. RELXILL allows $\log\xi$ to vary between 0 (neutral) and 4.7 (highly ionized). Our best-fit $\log\xi=1.7$, corresponding to a moderately ionized accretion disk. The $P_\mathrm{rej}$ is 98.2\%. RELXILL somewhat improves the fit quality, predicting a softer spectrum than BORUS, but it is still unacceptable considering the high $P_\mathrm{rej}$ and similar residuals to those in Figure \ref{fig:figure1a} without a clear systematic trend. While the RELXILL model is still likely not optimal for modeling the reflection from a TDO, our basic tests with it indicate that ionized reflection is unlikely to provide a satisfactory fit for the 2013 spectrum.

In 2023, we note J1521+5202 is only \hbox{X-ray} weak by a factor of $\approx 4$
and has a softer observed continuum shape (see Table~\ref{tab_obs}). Thus, Compton-thick absorption and Compton reflection are unlikely to be relevant in 2023, and
this is confirmed by our spectral-fitting tests.

\subsection{Spectral Energy Distribution and Light Curves}\label{sec:SEDlc}
\begin{figure*}[htbp]
  \centering
  \includegraphics[width=\textwidth]{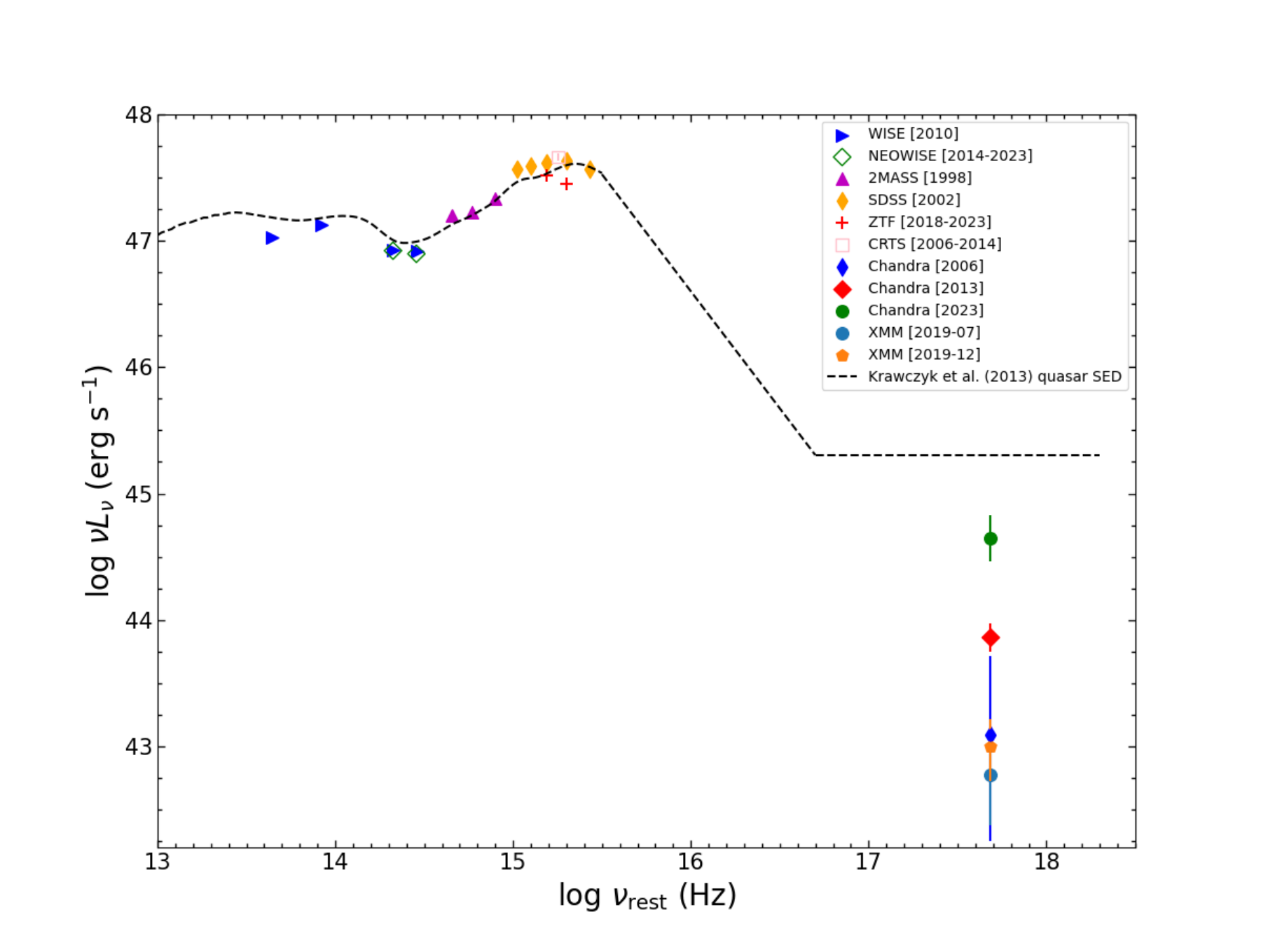}
  \caption{ IR-to-\hbox{X-ray} SED for J1521+5202.  IR-to-optical data points are obtained from the WISE, NEOWISE, 2MASS, ZTF, and CRTS
catalogs, where the NEOWISE, ZTF, and CRTS data points show average measurements of multi-epoch observations. The rest-frame 2 keV luminosities are taken from Table~\ref{tab_obs}. The dashed curve shows the mean luminous quasar SED from \citet{Krawczyk+2013} normalized to the 2500 \AA\ luminosity; the \hbox{X-ray} component is a  $\Gamma=2$ power-law continuum with 2 keV luminosity determined from the $\alpha_{\rm OX}\textrm{--}L_{\rm 2500 \AA}$ relation in \citet{Just+2007} ($f_\mathrm{weak}$= 0).\label{fig:sed}}
\end{figure*}

In Figure~\ref{fig:sed}, we show the IR-to-\hbox{X-ray} SED for J1521+5202. The \hbox{IR-optical} measurements are collected from the Wide-field Infrared Survey Explorer \citep[WISE;][]{WISE+2010}, \hbox{Near-Earth} Object WISE Reactivation \citep[NEOWISE;][]{NEOWISE2011}, Sloan Digital Sky Survey \citep[SDSS;][]{2000SDSS}, Zwicky Transient Facility  \citep[ZTF;][]{ZTF2019}, and Catalina \hbox{Real-Time} Transient Survey \citep[CRTS;][]{2009CRTS}. We plotted the average measurements of multi-epoch observations for the NEOWISE, ZTF, and CRTS data. All SED data and light curves have been corrected for Galactic extinction using the extinction law in \citet{2019extinction} with $R_V=3.1$. The Galactic $E(B-V)$ value of J1521+5202 was 0.019 \citep{1998extinction}. We did not present observed UV data for J1521+5202 because the strong intergalactic medium absorption made these points uncertain. We added the 2 keV luminosity determined from the best-fit results in Table~\ref{tab_obs}. For comparison, we included the mean SED of high-luminosity quasars in \citet{Krawczyk+2013}, scaled to the 2500~\AA~luminosity.  The \hbox{IR–optical} SED of J1521+5202 was broadly consistent with those of typical luminous quasars.  However, this quasar showed remarkable \hbox{X-ray} weakness in most epochs and \hbox{X-ray} variability.  \par
\begin{figure*}[htbp]
  \centering
  \includegraphics[width=\textwidth,height=8in]{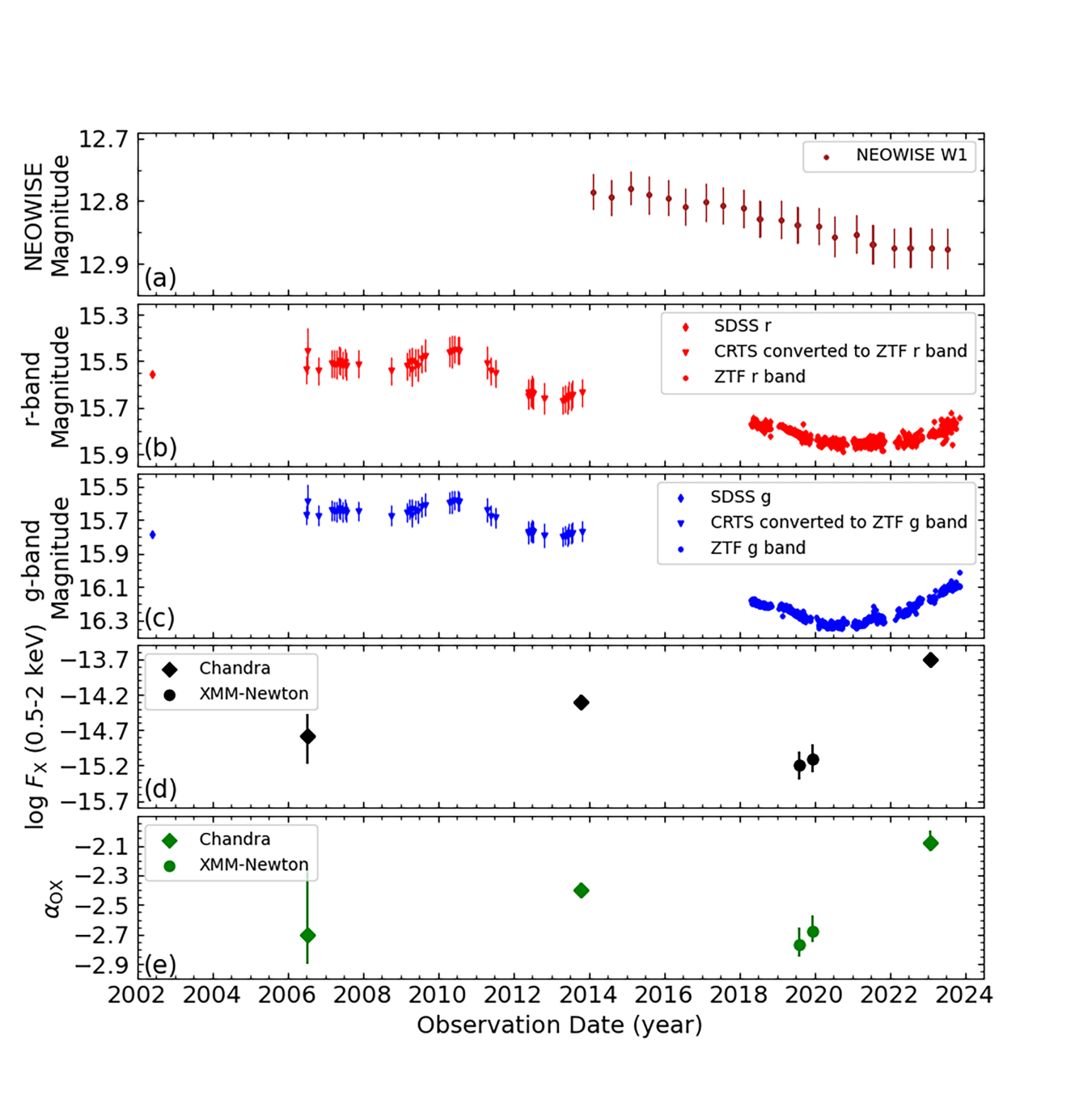}
  \caption{Light curves for the (a) NEOWISE W1 band (3.4 $\mu$m), (b) ZTF $r$-band, (c) ZTF $g$-band, (d)  observed-frame \hbox{0.5–2 keV} band (note the logarithmic scaling), and (e) $\alpha_\mathrm{OX}$. We group intraday measurements and convert the CRTS magnitude to ZTF $g$-band and $r$-band magnitudes using the filter transmission curves and SDSS spectrum.\label{fig:lightcurve}}
\end{figure*}

The IR-optical light curves are shown in panels (a), (b), and (c) of Figure~\ref{fig:lightcurve}, obtained from the \hbox{NEOWISE}, ZTF, and CRTS catalogs. The observed-frame \hbox{0.5--2~keV} flux ($F_\mathrm{X}$) and $\alpha_\mathrm{OX}$ derived from Table~\ref{tab_obs} are shown in panels (d) and (e), respectively. For the \hbox{IR–optical} light curves, we grouped any intraday measurements. We converted the CRTS magnitudes to ZTF $g$- and \hbox{$r$-band} magnitudes using the filter transmission and SDSS spectrum. We also computed the conversion uncertainty from CRTS to ZTF. The uncertainty caused by the deviation of the optical continuum power-law index was 0.01 mag in the $r$-band and 0.02 mag in the $g$-band. The cross-instrument uncertainty was 0.24 mag in the $r$-band and 0.25 mag in the $g$-band. In panels (b) and (c), we also included the SDSS $g$- and $r$-band photometric measurements. In panel (a), the maximum variability amplitude in the W1 band (3.4~$\mu$m) reached $\approx$~0.1 mag ($\approx$~10\%). In panels (b) and (c), the maximum long-term optical variability amplitudes reached $\approx$~0.4 mag ($\approx$~45\%) in the $r$-band and $\approx$~0.8 mag  ($\approx$~100\%) in the $g$-band. In panel (d), the largest variability amplitude of $F_\mathrm{X}$ was observed between the 2019 XMM-Newton observation and the 2023 Chandra observation and reached a factor of $32^{+31}_{-16}$, showing strong \hbox{X-ray} variability in 0.97 rest-frame years. Such large long-term \hbox{X-ray} variability amplitudes are rare among quasars \citep[e.g.,][]{Yang+2016,Middei+2017,Timlin+2020}. Following Section 4.2 of \citet{Timlin+2020}, this variation was a $\approx 7.6~\sigma_\mathrm{MAD}$ event, where $\sigma_\mathrm{MAD}$ is a robust estimator of the standard deviation derived using the median absolute deviation (MAD).
\section{Summary and future work}
We have reported on the \hbox{X-ray} spectra and variability of the optically ultraluminous WLQ J1521+5202 observed by Chandra and XMM-Newton. The key observational findings are the following:
\begin{enumerate}
\item J1521+5202 shows remarkable \hbox{X-ray} weakness and small $\Gamma_\mathrm{eff}$ values in all \hbox{X-ray} observations spanning 17 years, indicating consistently strong \hbox{X-ray} absorption. See Section~\ref{sec:X_analy}.
\item The 2023 Chandra spectrum can be acceptably described by a power law modified by heavy intrinsic absorption, but this model does not fit the 2013 spectrum well. See Sections~\ref{sec:X_analy} and~\ref{sec:testmodels}.
\item J1521+5202 has a typical quasar IR-to-optical SED, and the \hbox{long-term} IR/optical variability amplitudes are mild (e.g., a maximum variability amplitude of $\approx$ 45\% in the $r$-band). However, this quasar shows remarkable concurrent \hbox{X-ray} flux and spectral variability. For example, the observed-frame \hbox{0.5--2 keV} flux varies by a factor of $32^{+31}_{-16}$ in 0.97 rest-frame years between the 2019 XMM-Newton observation and the 2023 Chandra observation. See Section~\ref{sec:SEDlc}.
 \end{enumerate}

The overall \hbox{X-ray} and multiwavelength properties of J1521+5202 are qualitatively consistent with expectations for the TDO model, where the large \hbox{X-ray} variations could be driven by changes in the column density and/or the covering factor of the TDO, as the wind is a highly dynamical structure that would likely change on the observed multi-year timescales substantially. Besides, optical/UV photons produced on larger scales in the accretion disk remain largely unobscured  \citep[see Figure 1 of][]{Ni+2018}, and one would likely expect modest IR/optical variability much like that of a typical quasar, which is consistent with our IR-optical SED and light curves (Figure~\ref{fig:lightcurve}). In this scenario, the line-of-sight \hbox{X-ray} absorption by the TDO would be strong during the 2013 Chandra observation and even stronger during the 2019 XMM-Newton observations, and then it would drop greatly by the time of the 2023 Chandra observation. The intrinsically nominal level of \hbox{X-ray} emission relative to the optical/UV emission inferred from the 2023 Chandra spectrum (Section~\ref{sec:X_analy}) supports this scenario. Unfortunately, the other \hbox{X-ray} spectra do not allow us to determine reliably any absorption parameters (see Sections~\ref{sec:X_analy} and \ref{sec:testmodels}), and it is not clear whether it was variations of the TDO column density or TDO covering factor that drove the \hbox{X-ray} variability. The fact that a Compton-reflection model does not fit the 2013 Chandra spectrum well indicates that the thick disk itself, which is expected to be highly Compton thick, is probably not lying along our line-of-sight. Rather, at least in the case of J1521+5202, the outflow from the thick disk is probably what lies along our line-of-sight.
 
The general similarity of the \hbox{X-ray} variability of J1521+5202 to that of other WLQs with apparent TDO variations, including PHL~1092 \citep{Minuutti+2012}, SDSS~J1539+3954 \citep{2020Ni}, and SDSS J1350+2618 \citep{Liu+2022}, also supports an absorption-variability interpretation. The extensively studied low-redshift ($z=0.18$) quasar PDS\,456 also has C~IV properties similar to those of WLQs \citep[e.g.,][]{2005brien} and shows impressive \hbox{X-ray} absorption and luminosity changes \citep[e.g.,][]{2020reeves,2021reeves}.

The remarkable \hbox{X-ray} spectral and variability properties of J1521+5202 and \hbox{X-ray} weak WLQs generally require more photons per epoch for detailed study. This would require expensive Chandra or XMM-Newton observations (hundreds of ks per epoch). Furthermore, these observations would have risk owing to the strong \hbox{X-ray} variability of J1521+5202; note this strong variability already negatively impacted the 161~ks of XMM-Newton observations in 2021. Next-generation \hbox{X-ray} observatories, including Athena  \citep{2013Athena} and Lynx \citep{2019Ly}, will have photon-collecting areas at least an order-of-magnitude larger than for Chandra, and thus would allow much more efficient spectroscopic and variability investigations of J1521+5202 and related WLQs. If Athena and Lynx can maintain flexible observation scheduling, then a short observation could be obtained to check the flux level efficiently and decide if a proximate longer spectroscopic observation is merited. 

\vspace{5mm}
SW acknowledges financial support from Nanjing University. WNB, ZY, and FZ acknowledge financial support from Chandra \hbox{X-ray} Center grant GO2-23083X, the Penn State Eberly Endowment, and Penn State ACIS Instrument Team Contract SV4-74018 (issued by the Chandra \hbox{X-ray} Center, which is operated by the Smithsonian Astrophysical Observatory for and on behalf of NASA under contract NAS8-03060). BL acknowledges financial support from the National Natural Science Foundation of China grant 11991053. The Chandra ACIS Team Guaranteed Time Observations (GTO) utilized were selected by the ACIS Instrument Principal Investigator, Gordon P. Garmire, currently of the Huntingdon Institute for \hbox{X-ray} Astronomy, LLC, which is under contract to the Smithsonian Astrophysical Observatory via Contract SV2-82024.

\bibliography{sample631}
\bibliographystyle{aasjournal}
\end{document}